**Topological surface states interacting with bulk excitations in the Kondo insulator SmB$_6$ revealed via planar tunneling spectroscopy**

Wan Kyu Park[a,1], Lunan Sun[a], Alexander Noddings[a], Dae-Jeong Kim[b], Zachary Fisk[b], and Laura H. Greene[a,1,2]

[a]Department of Physics and Materials Research Laboratory, University of Illinois at Urbana-Champaign, Urbana, IL 61801; and [b]Department of Physics and Astronomy, University of California, Irvine, CA 92697

[1]To whom correspondence may be addressed. Email: wkpark@illinois.edu or lhgreene@magnet.fsu.edu.

[2]Present address: National High Magnetic Field Laboratory and Department of Physics, Florida State University, Tallahassee, FL 32310.

## Significance

Topological insulators are an emerging class of quantum matter in which the nontrivial bulk band structure gives rise to topologically protected surface states. Most are weakly correlated band insulators such as $Bi_2Se_3$. Despite numerous investigations on the strongly correlated bulk Kondo insulator samarium hexaboride ($SmB_6$), the topological nature of its surface states remained a mystery. Planar tunneling spectroscopy adopted in this work not only reveals spectroscopic signatures for the Dirac fermion surface states, but also reveals that their topological protection is not as robust as in weakly correlated systems due to interaction with collective bulk spin excitations. This finding suggests important implications on generic topological materials whose ground states are governed by strong correlations.

**Samarium hexaboride ($SmB_6$), a well-known Kondo insulator in which the insulating bulk arises from strong electron correlations, has recently attracted great attention owing to increasing evidence for its topological nature, thereby harboring protected surface states. However, corroborative spectroscopic evidence is still lacking, unlike in the weakly correlated counterparts, including $Bi_2Se_3$. Here we report results from planar tunneling that unveil the detailed spectroscopic properties of $SmB_6$. The tunneling conductance obtained on the (001) and (011) single crystal surfaces reveal linear density of states as expected for two and one Dirac cone(s), respectively. Quite remarkably, it is found that these topological states are not protected completely within the bulk hybridization gap. A phenomenological model of the tunneling process invoking interaction of the surface states with bulk excitations (spin excitons), as predicted by a recent theory, provides a consistent explanation for all of the observed features. Our spectroscopic study supports and explains the proposed picture of the incompletely protected surface states in this topological Kondo insulator $SmB_6$.**

Topological insulators are an emerging class of quantum matter in which the nontrivial topology of the bulk band structure naturally gives rise to topologically protected, *i.e.*, robust, surface states (1, 2). Several dozens of such materials have been discovered but most of them, including $Bi_2Se_3$, are weakly correlated band insulators. A recent theoretical proposal (3) that certain Kondo insulators (4), which are insulating in the bulk due to strong electron correlations, could also be topological has stimulated vigorous research in the field. In particular, samarium hexaboride ($SmB_6$) has drawn great attention owing to its telltale resistivity behavior saturating below 4 K (5). Various experiments have been implemented to investigate this possibility (6-24), establishing the robustness of the surface states and the Kondo hybridization leading to a formation of the bulk gap, but their topological origin and nature has not been unambiguously confirmed. Factors contributing to this situation include their inherently complex nature due to strong correlations, non-trivial surface chemistry, and insufficient energy resolution. A recent report of quantum oscillations in magnetic torque supports the topological origin of the surface states (22), but conflicting results have also been reported (23).

Planar tunneling spectroscopy, inherently surface-sensitive with high energy resolution and momentum selectivity, is ideally suited for the study of surface states, particularly so in $SmB_6$ where the bulk hybridization gap is much smaller than the band gap in $Bi_2Se_3$. Lead (Pb) is chosen as the counter-electrode in this study because the quality of its measured superconducting density of states (DOS) is an important junction diagnostic and, as we will show, this choice is of crucial importance in unveiling the nature of the surface states. As shown in Fig. 1*A*, the sharpness of Pb superconducting features in the differential conductance, $g(V) = dI/dV$, confirms the high quality of the junctions. Spectroscopic properties of $SmB_6$ are revealed more clearly when the Pb is driven normal by temperature or applied magnetic field. The $g(V)$ curves from both (001) and (011) surfaces show a peak at −21 mV, arising from the bulk hybridization gap, in agreement with angle-resolved photoemission spectroscopy (ARPES) results (13). Note, in particular, both surfaces exhibit linear conductance at low bias with a V-shape around a minimum slightly below zero bias, as expected for Dirac fermion DOS. Quite notably, the linearity ends at ~4 mV, well below the bulk gap edge. As detailed below, a careful analysis of this behavior, taken with that when the Pb is superconducting, leads to unraveling the intriguing topological nature of the surface states in $SmB_6$.

Bulk $SmB_6$ is insulating because the chemical potential falls within the gap arising from Kondo hybridization of the itinerant 5*d* bands with the localized 4*f* bands. The resulting hybridization gap is substantially reduced due to the inherent strong correlations (4). Colored-contour maps of the normalized conductance over the temperature range of 1.72 K – 100 K (Figs. 1*B* and *C*), show that there are several distinct stages in the temperature evolution; also clearly shown in the normalized conductance curves taken at fixed bias voltages (Fig. 1*D*). As the temperature is lowered, the hybridization begins to appear at 70 K – 80 K but evolves slowly due to thermal broadening and valence fluctuations (25). Below 48 K – 53 K, signatures for gap formation appear, including a distinct peak at −21 mV and a rapid suppression around zero bias. Theoretically, topological surface states should exist concomitantly with the bulk gap opening. But here, no clear evidence of the surface states is detected down to 25 K – 30 K. Below this, the conductance taken at +4 mV decreases more slowly due to some contribution from the surface states and below 15 K – 20 K the conductance increases, indicative of a stronger surface state contribution. Note this behavior corresponds with the resistivity data, which shows a hump in the logarithmic plots and subsequent decrease of the slope (*SI Appendix*, section 2). Furthermore, below 5 K – 6 K, the conductance at fixed bias exhibits an abrupt increase followed by saturation in both orientations, although weaker on the (011) surface. This is in accord with the resistivity saturating below 4 K due to the dominance of the surface states. Being sensitive to surface DOS, tunneling spectroscopy offers more detailed information on their temperature evolution.

The V-shaped DOS expected for surface Dirac fermions is manifested as linear conductance at low bias (Fig. 2*A*). A noticeable difference between the two surfaces is that the (001) surface exhibits two distinct slopes while only one is seen in the (011) surface. We point out that if the origin of the surface states in $SmB_6$ were trivial (16), e.g., arising from impurities (26), the tunneling conductance would show a peak at the corresponding impurity band energy, but no such features are seen in our data. The DOS of Dirac fermions is inversely proportional to the Fermi velocity squared (*SI Appendix*, section 8). Considering the discrepancy in the Fermi velocities reported from different measurements (10, 13, 14, 22), it would be valuable to extract Fermi velocities from tunneling data, e.g., conductance oscillations due to Landau level quantization under magnetic field. But as seen in Fig. 1*A*, such oscillations are not observed up to 9 T. And, since the tunneling matrix element that appears in the coefficient of the conductance formula is not known, it is not possible to extract an absolute value of the Fermi velocity by simply taking the conductance slope. However, we can compare the slopes to obtain the ratio of the Fermi velocities for different Dirac fermions residing on a given surface (*SI Appendix*, section 8). Thus, we decompose the linear conductance region into contributions from one (or two) Dirac cone(s) (Fig. 2*B*). In turn, the two Dirac bands on the (001) surface are identified as the α and γ bands by comparing the slopes of the decomposed linear conductance with the reported Fermi velocities (22). Likewise, the linear conductance on the (011) surface originates from the single (β) Dirac band. Theories (3, 27-30) predict that topological surface states exist around the projected $\overline{X}$ points in the surface Brillouin zone. They correspond to the $\overline{\Gamma}$ and $\overline{X}$ points on the (001) surface and the $\overline{Y}$ point on the (011) surface. Thus, as summarized in Fig. 2*C*, the association of the topological surface states as described above is in good agreement with theory; and also with results from quantum oscillations (22) and ARPES measurements (13, 14, 18). The two Dirac points on the (001) surface are at −0.4 meV (α band, electron-like) and +2.3 meV (γ band, hole-like). The Dirac point on the (011) surface is at −0.2 meV (β band, electron-like). These small values indicate that the Dirac points are located close to the chemical potential, i.e., well inside the bulk gap region, in contrast with some ARPES results (14). This discrepancy may result from the chemical potential being sensitive to the surface chemistry. More specifically, while the cleaved surfaces measured by ARPES are likely to be electrically non-neutral due to dangling bonds, as shown experimentally (16, 20, 21), we argue that they are pacified in our junctions via the surface oxidation process to form a tunnel barrier (*SI Appendix*, section 2). That the surface states are still detected after the harsh process of polishing and oxidation attests to their robustness and thus, their topological origin. Also, that these states are moved to beneath the tunnel barrier oxide layer we form is consistent with recent ion damage experiments (9).

As noted earlier, the linearity in conductance ends at a relatively low bias, where $g(V)$ shows a kink near +4 mV or a broad hump around −4 mV. Here, we show how this behavior points to the non-trivial nature of the topological surface states. First, a detailed scrutiny of the kink-hump structure shows that the characteristic voltages are nearly temperature independent (*SI Appendix*, section 4). Second, the 4 mV kink appearing when the Pb is in the normal state becomes a pronounced peak as the Pb becomes superconducting and moves to a higher bias with decreasing temperature (Fig. 3*A*), suggesting that understanding the features in the superconducting state may hold the key. For a detailed analysis, conductance curves in the superconducting state are normalized against those obtained when the Pb is driven normal by the applied magnetic field of 0.1 T (Fig. 3*B*). With Pb in the superconducting state at the lowest temperature, the two DOS coherence peaks (indicated as V− and V+) are clearly visible. But, their temperature evolution is quite unusual: With increasing temperature, the conductance at the negative bias coherence peak (V−) becomes larger than the positive bias counterpart (V+). There is also an additional peak outside the gap, which appears only in the positive bias branch at $V_{1(2)}$ on each surface. Although the $V_{1(2)}$ peak energy is close to that of the phonons in Pb ($\omega_{ph}$), its phononic origin can be ruled out for two reasons (*SI Appendix*, section 5). First, the Pb phonon features are seen in the tunneling conductance because they are embedded in the superconducting DOS via strong electron-phonon coupling (bulk physics) (31), and thus appear symmetrically in bias. Second, the peaky conductance shape at $V_{1(2)}$ is different from the hump-dip shape of the phonon features; The phonons only appear as peaks in second harmonic measurements (31). The asymmetric appearance of such a pronounced peak and the asymmetric temperature evolution of the coherence peaks rule out the phonons in the Pb as the origin. Inelastic tunneling processes through sources that are extrinsic to the two electrodes and exist in or near the tunnel barrier would appear as kinks (not peaks) in the conductance at *symmetric* bias voltages, so are also ruled out (*SI Appendix*, section 11). The $V_{1(2)}$ peak changes to a kink when the Pb is driven normal either by temperature ($T_c$ = 7.2 K; $H$ = 0 T) or by applied magnetic field (1.72 K; 0.1 T), as seen in Figs. 3*A* and 1*A*, respectively. These observations indicate that the peak at $V_{1(2)}$ and the kink at $\omega_{1(2)}/e$ (Fig. 2*A*) have the same origin. In Fig. 3*C*, we show the temperature evolution of the peaks at V− and $V_{1(2)}$ quantitatively. On the left, we plot the central positions of the kinks plus superconducting gap values, $(\Delta+\omega_{1(2)})/e$, along with the peak positions, $V_{1(2)}$. Note their temperature dependences track closely below $T_c$, indicating their connection to the superconducting DOS. On the right, we plot the relative peak heights of the negative bias coherence peak (V−) and the $V_{1(2)}$ peak with respect to the positive bias coherence peak (V+). Both peaks' heights increase with increasing temperature quite rapidly.

To explain these exotic features, we invoke an inelastic tunneling model (*SI Appendix*, section 11) involving bosonic excitations of energy $\omega = \omega_0$ (= $\omega_1$ or $\omega_2$) in the SmB$_6$. Figure 3*D* depicts two particular bias configurations. First, consider the left panel. When $eV = \Delta+\omega_0$, the tunneling probability is enhanced due to additional channels opened via *emission* of such bosons. This inelastic tunneling through emission of bosons requires the electrons tunneling from the Pb to have a minimum excess energy of $\omega_0$ when arriving at the SmB$_6$, which is dissipated in the emission process. And its contribution shows up as a pronounced peak at $V_{1(2)}$, where the tunneling electrons originate from the peaky DOS of Pb (*i.e.*, the coherence peak at −Δ). This also naturally explains why the $V_{1(2)}$ peak position follows $(\Delta+\omega_{1(2)})/e$ with temperature and then changes to a kink at $\omega_{1(2)}/e$ when the Pb is driven normal (flat DOS and Δ = 0). Furthermore, this is consistent with the relative height of the $V_{1(2)}$ peak growing with temperature: With increasing temperature, the Pb superconducting gap closes and thermal population effects broaden the coherence peaks. Such broadening effects are less severe in the inelastic tunneling at $V_{1(2)}$ because the associated energy has a constant value ($\omega_{1(2)} \sim$ 4 meV) that is much larger than the energy corresponding to the V+ peak, namely, $\Delta(T)$, which has a maximum value of 1.4 meV and decreases with increasing temperature. Now, consider Fig. 3*D*, *Right*. When the SmB$_6$ is biased negatively, inelastic tunneling involves only *absorption* of bosons: Surface state electrons at deep energy levels can pop up into the chemical potential by absorbing bosons and thus can participate in the tunneling. Unlike in the emission process, the conductance contribution from this absorption process depends on the population of bosons ($n_\omega$) in the $SmB_6$ and that follows the Bose-Einstein distribution, $n_\omega=1/[exp(\omega/k_BT)-1]$. This contribution is maximized when $eV = -\Delta$ because of the large and peaked empty DOS of the superconducting Pb into which the electrons from the $SmB_6$ tunnel at this bias. As the bias is moved away from this coherence peak, the inelastic tunneling is concomitantly suppressed, so its contribution is not as pronounced as at $eV = -\Delta$. This explains why it is noticeable only at V− in the conductance data. The inelastic contribution at this bias follows from the Bose-Einstein statistics: Since there is an exponential increase in $n_\omega$ with increasing temperature, the relative height of the V− peak must increase rapidly with increasing temperature, as seen in Fig. 3*C*. As described above, since the inelastic processes in the $SmB_6$ occur asymmetrically with respect to the bias voltage, the asymmetric features observed in the tunneling conductance are consistently explained. The same processes but involving bosonic excitations (phonons) in the Pb instead of the $SmB_6$ should occur at reversed biases and they cannot explain the experimental features (*SI Appendix*, section 11).

We have shown how inelastic tunneling via emission and absorption of bosonic excitations in $SmB_6$ can explain all the features observed when the counter-electrode is in the superconducting or normal state. Low energy phonon modes in $SmB_6$ are detected at 2.6 meV and 11.6 meV (32), well separated from the $\omega_{1(2)}$ (~4 meV) value detected in our tunneling spectroscopy, so they can be ruled out as the responsible excitations. Recent inelastic neutron scattering measurements (24) reported a strong resonance mode at 14 meV identified as spin exciton excitations in the bulk, implying that $SmB_6$ is not too far from an antiferromagnetic quantum critical point. Building on this finding, a recent theory (33) proposed that the topological protection of the surface states in $SmB_6$ is incomplete due to their strong interaction with the bulk spin excitons. The features due to such interaction are then detected in usual elastic tunneling since they are embedded in the spectral density of the surface states via self-energy corrections due to emission and absorption of virtual spin excitons, similarly to the case of Pb phonon features observed in the tunneling conductance when Pb is superconducting. In our tunneling conductance spectra, they show up as a kink and a hump at $\pm\omega_{1(2)}$, respectively, when the Pb is driven normal. The asymmetric appearance and evolution of the $V_{1(2)}$ and V− peaks are due to additional tunneling channels opened up via inelastic processes and their contributions are more pronounced when the Pb is superconducting due to the sharpened DOS.

Our tunneling model involving the spin excitons in $SmB_6$ can explain the features in both normal and superconducting states. Based on this result, we argue that the hump slightly below the chemical potential observed in an ARPES study (13) has the same origin as that in our tunneling conductance, that is, absorption of virtual spin excitons by the surface states (*SI Appendix*, section 10). The spin exciton energy ($\omega_{1(2)}$) detected in our measurements is much smaller than the bulk value observed in inelastic neutron scattering measurements (24) because the antiferromagnetic coupling at the surface ($J'$) is reduced substantially, as is the spin exciton energy (33). The incomplete protection of the surface states in $SmB_6$ is in strong contrast to the case of weakly correlated topological insulators such as $Bi_2Se_3$, in which the surface states span the entire region within the bulk gap. It is noteworthy that the decisive influence of spin excitons on the topological surface states in $SmB_6$ is rooted on the bulk physics involving strong correlations since they are collective excitations in the bulk, suggesting that similar possibilities should be taken into account in the study of other strongly correlated topological insulators.

Processes involving spin excitons interacting with the surface states diminish the protection provided by their topological nature. This, in turn, affects the temperature evolution of the surface states, explaining why the decrease in conductance at 4 mV starts to slow down only below 25 K – 30 K (Fig. 1*D*) instead of the higher bulk gap opening temperature. The rapid conductance jump below 5 K – 6 K and subsequent saturation indicate that protected (or coherent) low-energy surface states may exist only in this low temperature region as the interaction with spin excitons becomes negligible. Similar abrupt changes in several other experiments (34, 35) could also be understood by following this reasoning. A more deepened microscopic understanding of the tunneling process, particularly regarding the relevant length scales (33, 36) and their temperature evolution (*SI Appendix*, section 12), is expected to facilitate progress toward a more comprehensive picture of the topological states arising in strongly correlated electron systems.

**Materials and Methods**

High-quality $SmB_6$ single crystals are grown by a flux method. For tunnel junction fabrication, crystals with well-defined facets are embedded in epoxy molds and polished down to sub-nm roughness. The tunnel barrier is formed by plasma oxidation of the crystal surface in a high vacuum chamber and the counter-electrode is thermally evaporated through a shadow mask. Differential conductance is measured using standard four-probe lock-in technique in a Quantum Design PPMS Dynacool system. For further details, *SI Appendix*, section 1.

**ACKNOWLEDGMENTS.** W.K.P. thanks M. Dzero and P. Riseborough for fruitful discussions and P. Riseborough also for the help with simulation of the surface spectral density, C. O. Ascencio and R. Haasch for the help with surface analyses. A.N. is supported by John A. Gardner Undergraduate Research Award. This material is based upon work supported by the US National Science Foundation (NSF), Division of Materials Research (DMR) under Award 12-06766, through the Materials Research Laboratory at the University of Illinois at Urbana-Champaign. The work done at UC-Irvine was supported by the US NSF DMR under Award 08-01253.

**FIGURE LEGENDS**

**Fig. 1.** Planar tunneling spectroscopy of SmB$_6$. (*A*) Typical $g(V)$ curves for the (001) and (011) surfaces taken at 1.72 K when the Pb is in the superconducting (H = 0.0 T) and normal states (H = 0.1 and 9.0 T). While high-bias features such as a peak at −21 mV overlap, large deviations are seen in the low bias region due to the superconducting gap opening in Pb. The overall conductance shape when the Pb is in the normal state, particularly the linearity at low bias, is essentially invariant under magnetic field up to 9 T. (*B* and *C*) Colored-contour maps of the conductance based on twenty three curves taken from 1.72 K to 100 K for each junction with the Pb kept in the normal state. Each $g(V,T)$ curve is normalized by dividing it out with the $g(V,100$ K) curve. The y-axis is in log scale for clear view. (*D*) Temperature dependence of $g(V)$ at a fixed bias of $V = -21$ mV and +4 mV, normalized by $g(-50$ mV). The x-axis is in log scale for clear view. From the inflection points in the $g(-21$ mV) curves, the bulk gap is inferred to open at 48 K – 53 K (upward arrows). The deviation of the $g(+4$ mV) curves from logarithmic dependence at 25 K – 30 K (downward arrows) signifies the contribution from the metallic surface states.

**Fig. 2.** Topological surface states in SmB$_6$. (*A*) $g(V) / g(-50$ mV) curves are linear at low bias, as depicted by the dashed lines, reflecting the V-shaped DOS for Dirac fermions. At the end of the linear region, they show a kink ($\omega_{1,2}/e \approx 4$ mV) and a broad hump (around −4 mV). The linear region for the (001) surface consists of two parts with different slopes, whereas the (011) surface exhibits only one slope. (*B*) For the (001) surface, added contributions from two Dirac cones with distinct Dirac points ($\varepsilon_D$) can reproduce the double linear conductance. For the (011) surface, the linear conductance arises from only one Dirac cone. Analysis of the slopes enables assigning them to corresponding surface bands ($\alpha$, $\beta$, and $\gamma$) in agreement with the literature (22). (*C*) Illustration of the surface Brillouin zone for each crystal surface and Fermi surfaces where the topological surface states are predicted to reside, in agreement with our results and other measurements in the literature.

**Fig. 3.** Interaction of the topological surface states with spin excitons. (*A*) Low-temperature $g(V)$ curves, shifted vertically for clarity. The asterisks are to show that the peak in the superconducting state evolves continuously, changing to a kink when the Pb is driven normal. (*B*) $g(V)$ curves in the superconducting state normalized against the normal state ($H = 0.1$ T) curves. The two peaks (V+ and V−) are the coherence peaks in the superconducting DOS of Pb. A well-pronounced additional peak is also observed at $V_{1(2)}$ but only in the positive bias branch. (*C*) Temperature evolution of the $V_{1(2)}$ and V− peaks. (*Left*) The $V_{1(2)}$ peak follows the temperature dependence of $(\Delta+\omega_{1,2})/e$, where $\Delta$ is the superconducting energy gap in Pb and $\omega_{1,2}/e$ denotes the kink position shown in Fig. 2A. (*Right*) Relative peak height to the V+ peak after subtracting the background value of 1. Both peaks at $V_{1(2)}$ and at V− increase rapidly with temperature. (*D*) Two bias configurations based on DOS diagrams for the superconducting Pb and the surface states in $SmB_6$ to account for the inelastic tunneling process. (*Left*) When $V = (\Delta+\omega_0)/e$, the excess energy ($\omega_0$) of tunneling electrons relative to the chemical potential ($\mu_R$) can open additional tunneling channels via emission of spin excitons (the *e-h* pair enclosed by an ellipse) in $SmB_6$. (*Right*) When $V = -\Delta/e$, the tunneling probability increases if electrons at energy $-\omega_0$ can also participate via absorption of spin excitons that are already available in the $SmB_6$. This absorption process contributes most greatly at $V = -\Delta/e$ around which the empty DOS in Pb shows a peak.

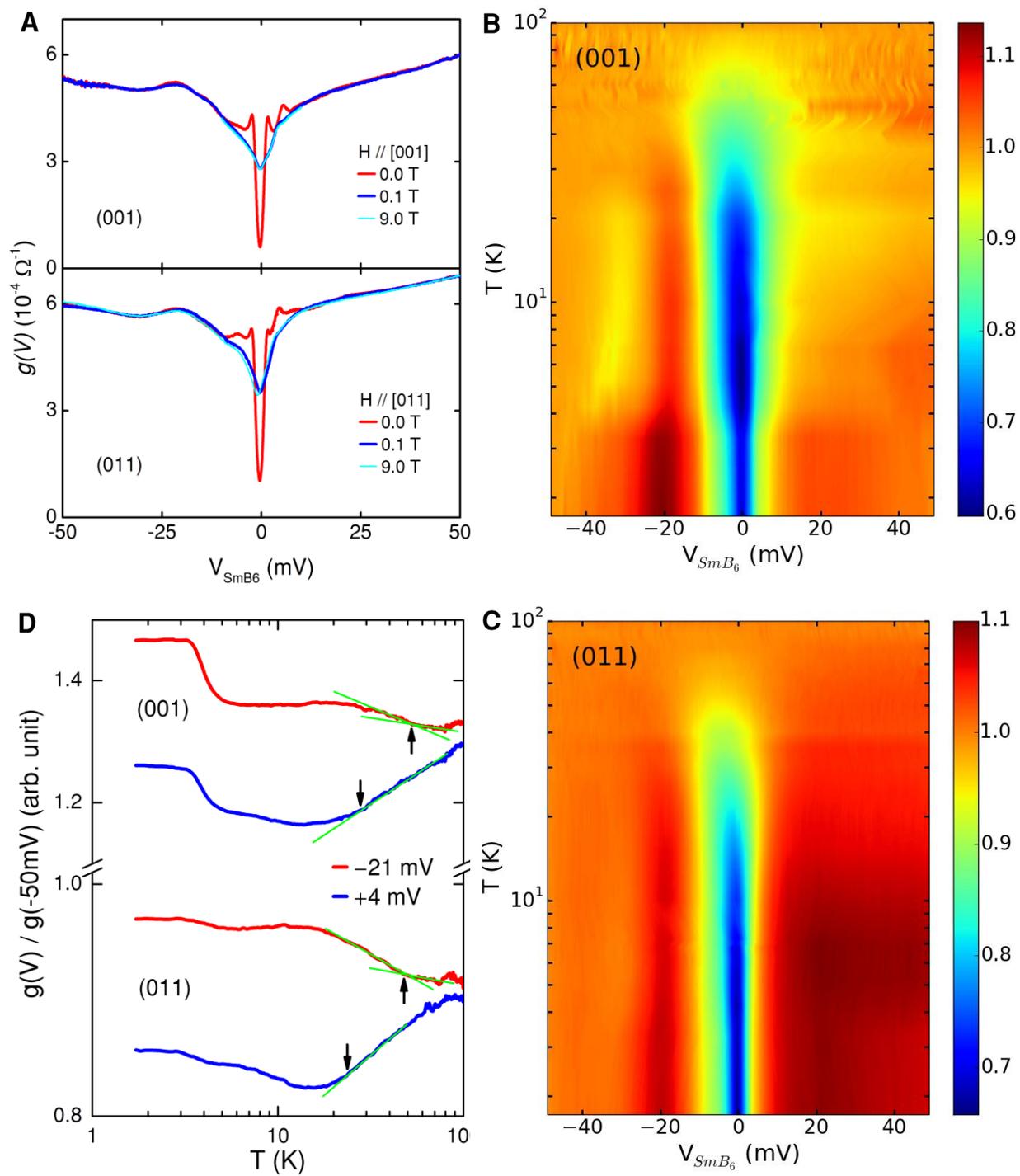

Fig. 1, W. K. Park *et al*.

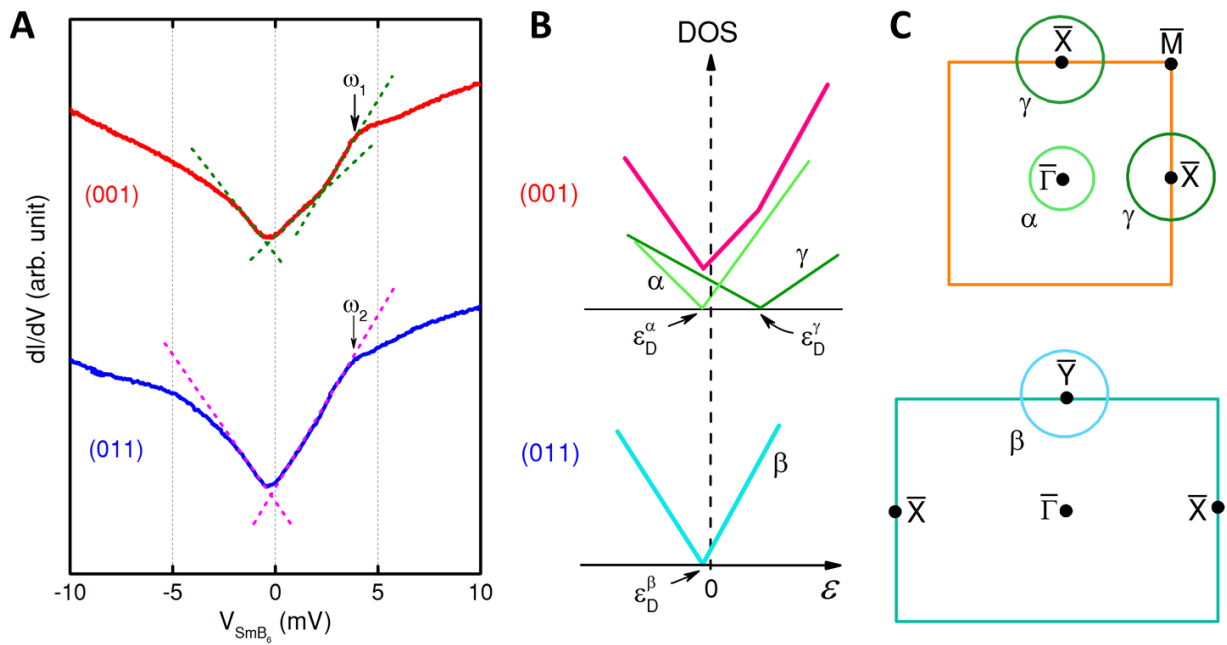

Fig. 2, W. K. Park *et al*.

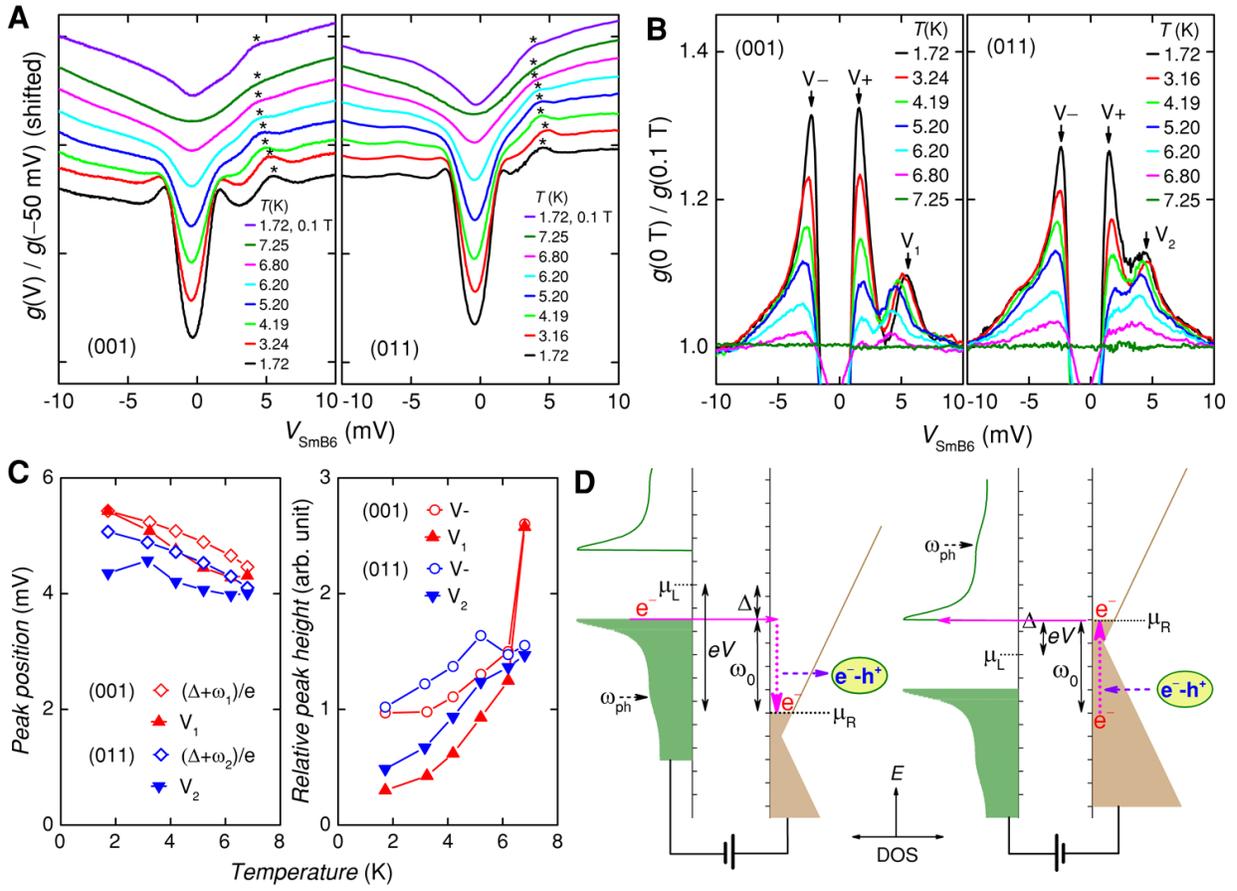

Fig. 3, W. K. Park *et al*.

# Supporting Information for

**Topological surface states interacting with bulk excitations in the Kondo insulator SmB$_6$ revealed via planar tunneling spectroscopy**

Wan Kyu Park[a,1], Lunan Sun[a], Alexander Noddings[a], Dae-Jeong Kim[b], Zachary Fisk[b], and Laura H. Greene[a,1,2]

[a]Department of Physics and Materials Research Laboratory, University of Illinois at Urbana-Champaign, Urbana, IL 61801; and [b]Department of Physics and Astronomy, University of California, Irvine, CA 92697

[1]To whom correspondence may be addressed. Email: wkpark@illinois.edu or lhgreene@magnet.fsu.edu

[2]Present address: National High Magnetic Field Laboratory and Department of Physics, Florida State University, Tallahassee, FL 32310

**This PDF file includes:**
    Materials and Methods
    Supporting Text
    Figures S1 to S13

## 1. Materials and Methods

SmB$_6$ single crystals were grown by a conventional flux method. As-grown crystals with flat surfaces of regular shape are chosen for the experiments in this work. Those with rectangular facets are used for the junctions on the (001) surface and hexagonal facets for the (011) surface since their shape is known to match with each surface orientation. For tunnel junctions, one crystal per each orientation has been used repeatedly more than fifty times by embedding in a mold made of Stycast® epoxy (2850-FT) and polishing. Since the surface roughness is a critical factor for a high-quality tunnel junction, the crystals are polished using alumina lapping films with particle size in the range of 12 – 0.3 μm. Our analysis of polished surfaces using an atomic force microscope (AFM) shows that they are extremely smooth with 0.4 – 0.8 nm peak-to-dip roughness. Polished crystals are loaded into a high-vacuum chamber for the formation of a tunnel barrier. A variety of methods have been attempted including sputter deposition of thin aluminum layers (1 – 7 nm thick) and subsequent oxidization using oxygen plasma generated by DC glow discharge. It is found that high-quality tunnel junctions can be produced most reproducibly by oxidizing the polished crystal surface. In order to remove any non-intrinsic chemical species that might exist on the crystal surface such as hydrocarbons and oxides, an argon ion beam etching is carried out just before the oxidization. The crystal edges are painted with thinned Duco® cement for insulation purpose prior to the deposition of the counter-electrode (Pb). Thin Pb strips are evaporated through a shadow mask. Figure S1 shows an optical image of a typical tunnel junction. Aluminum strips for contact leads are attached using a silver paint. If needed, DC current is ramped between the two leads contacting the crystal to reduce the contact resistance, whose typical value is less than a few ohms at room temperature. Differential conductance is measured using a standard four probe lock-in technique as a function of temperature and magnetic field.

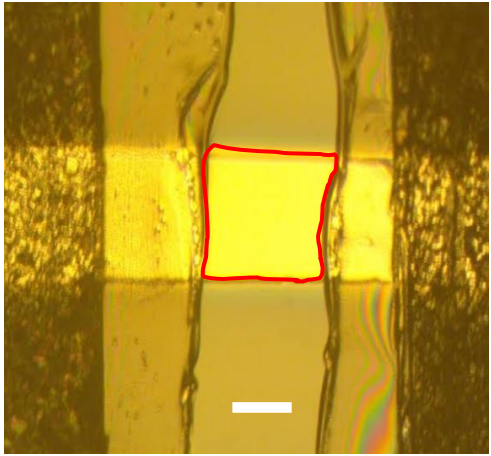

**Fig. S1.** Optical image of a typical tunnel junction. False color, top view. The SmB$_6$ crystal (bright area in the middle) is embedded in a Stycast® mold (dark areas outside) and polished to sub-nm roughness. The tunnel barrier is formed by oxidizing the crystal surface. The bright-colored horizontal strip is a Pb thin film evaporated as the counter-electrode. The white scale bar denotes 200 μm. The area enclosed by a red solid line denotes the junction area, whose size is typically ~0.1 mm$^2$.

## 2. Characterization of Plasma-Oxidized SmB$_6$ Crystal Surfaces

Electron spectroscopies including Auger and X-ray photoelectron spectroscopy (XPS) are employed for the analysis of surface chemistry. XPS data show a clear peak from oxidized boron atoms (B$_2$O$_3$) (1) on the plasma-oxidized crystal surfaces, suggesting that the surface B$_2$O$_3$ layer act as a tunnel barrier. This is a very likely possibility considering the large band gap (6 – 9 eV) of $B_2O_3$ as estimated from a band structure calculation (2). No clear evidence for oxidized samarium atoms is detected but the existence of samarium oxides cannot be ruled out completely considering the chemical complexity (3).

In order to see whether the crystals show similar transport properties before and after the plasma oxidation, DC resistance is measured on the same crystals used for tunnel junctions, polished and oxidized in the same manner. As displayed in Fig. S2, their temperature dependence remains almost the same, confirming that the surface states are still there even after the crystal surface is oxidized. Combined with the fact that the same oxidization process enables producing high-quality tunnel junctions, this observation attests to the robustness of the surface states in $SmB_6$. While further characterization using micro-structural analysis techniques such as secondary ion mass spectrometry (SIMS) is necessary to reveal the detailed depth profile for the chemical species, one can imagine that the surface states might have moved to beneath the top oxide layer. This speculation is very reminiscent of the results from transport measurement on ion-damaged $SmB_6$ crystals (4).

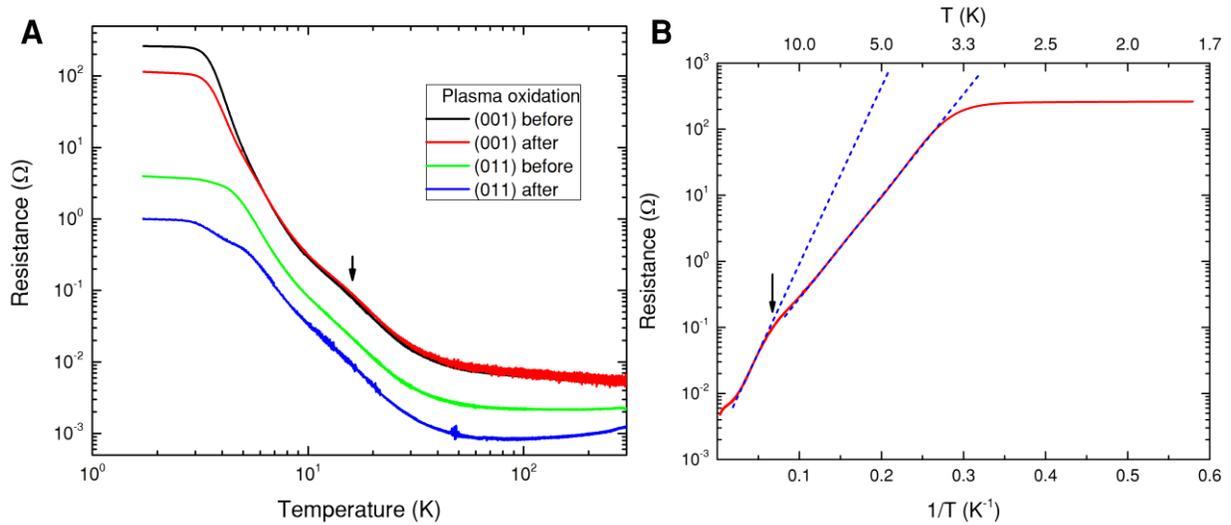

**Fig. S2.** Resistance vs. temperature of the $SmB_6$ crystals. (*A*) Resistance before and after the plasma oxidation plotted in log-log scale. (*B*) Resistance of the (001) crystal before the oxidation in log vs. 1/T plot. A slight difference in the magnitude arises from the difference in the contact geometry. Other than that, the temperature dependence remains almost the same. The downward arrows are to indicate a hump around 15 K – 20 K, a feature commonly observed in the literature to which not much attention has been paid. This may signify that the electrical conduction through the surface states has a sizable contribution to slow down the exponential increase due to the bulk gap formation (see the slope change around the hump as shown in *B*) but not strong enough to take it over completely till cooled down to 3 K – 4 K.

## 3. Magnetic Field Dependence

The tunnel junctions are measured under applied magnetic field up to 9 T. As shown in Fig. S3, essentially no changes are detected compared to the zero-field data. While this situation is in stark contrast to magnetic torque measurements (5) showing clear quantum oscillations, it is similar to the non-observation of oscillations in resistance measurements up to 45 T (see supplementary materials in Ref. 5). This implies that quantum oscillations in $SmB_6$ may work quite differently from those in $Bi_2Se_3$, in which quantum oscillations are clearly observed in all the three different kinds of measurements, namely, resistance, magnetic torque, and tunneling conductance (6-9). In fact, how Landau quantization works in strongly correlated systems like $SmB_6$ is an open question. A clue may be found in the fact that both tunneling conductance and electrical resistance reflect charge transport properties, whereas magnetic torque arises from magnetic moments.

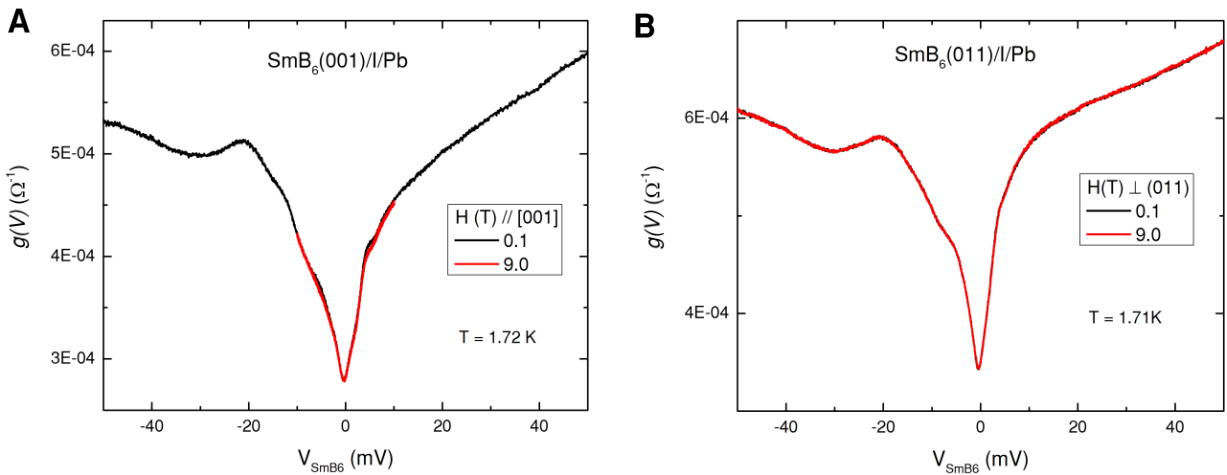

**Fig. S3.** Magnetic field dependence of tunneling conductance. Junctions on (*A*) the (001) surface, and (*B*) the (011) surface. Up to 9 T, essentially no field dependences are detected. This is the case whichever direction the field is oriented (perpendicular, 45 degrees or parallel to the crystal surface).

## 4. Temperature Evolution of Tunneling Conductance

Figure S4 shows the temperature evolution of tunneling conductance in line plots. The curves are normalized against the conductance at −50 mV but not against the one taken at 100 K. At high temperature, particularly in the (001) surface, fluctuations in the conductance are seen possibly due to thermal/valence fluctuations. Upon further lowering the temperature, features due to the bulk gap opening (peak at −21 mV) and the surface states (linearity at low bias) as described in the main text are showing up.
    Figure S5 shows temperature evolution of normalized conductance at several fixed bias voltages. At a low bias such as –0.4, 1, or 2 mV inside the kink-hump structure, the contribution from the surface states as signified by the slowdown below 20 K – 25 K (see text) shows up more slowly than at 4 mV since the surface density of states (DOS) is smaller whereas the impact of bulk gapping is larger at such biases.

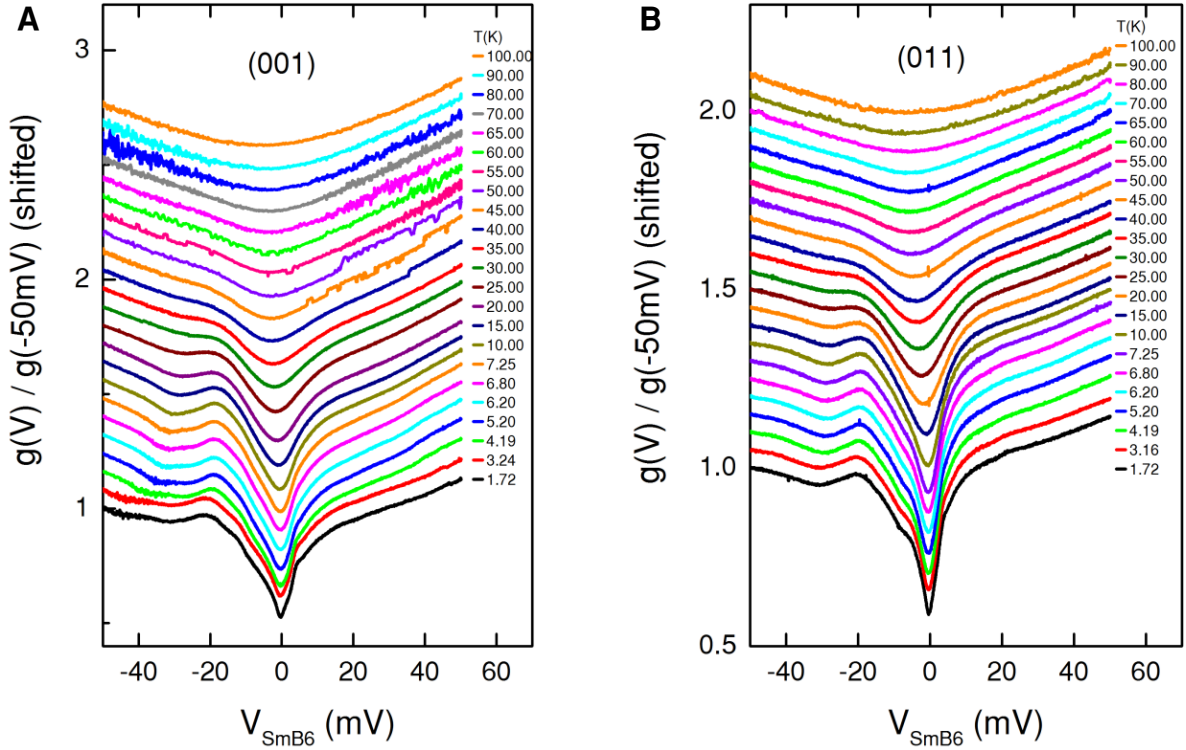

**Fig. S4.** Normalized conductance. Junctions on (*A*) the (001) surface and (*B*) the (011) surface. All curves except the 1.72 K one are shifted vertically for clarity. The colored contour plots shown in the main text (Fig. 1) are based on these data but normalized against the curve at 100 K.

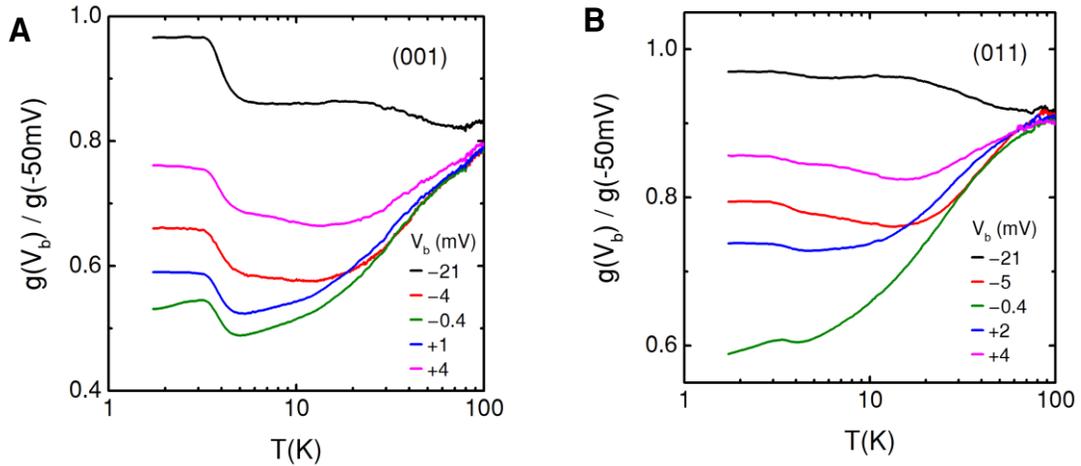

**Fig. S5.** Normalized conductance at a fixed bias. Junctions on (*A*) the (001) surface and (*B*) the (011) surface. More curves are plotted than in the main text. Other than the features described in the text, the rapid increase below 5 K – 6 K and subsequent saturation is clearly seen at all bias voltages except at the Dirac points (–0.4 mV). This common behavior at low temperature implies that the distinct change originates from some drastic evolution in the electronic properties of the system as a whole.

As described in the main text, the linearity in conductance ends around ±4 mV where it turns into a kink or a hump. Figure S6 presents zoomed-in plots to show their temperature evolution in detail. The kink-hump structure evolves continuously with the characteristic voltages remaining almost the same. By closely inspecting the conductance curves, one can roughly define at what temperature such features disappear. It is in the range of 15 K – 20 K, consistent with the range identified from Fig. 1d: the conductance turning from decrease to flattening then increasing as temperature is lowered, forming local minima. Presumably, this is the temperature below which some coherence in the surface states might begin to develop as the population of spin excitons decreases exponentially with temperature, resulting in a fully coherent (or protected) state below 3 K – 4 K.

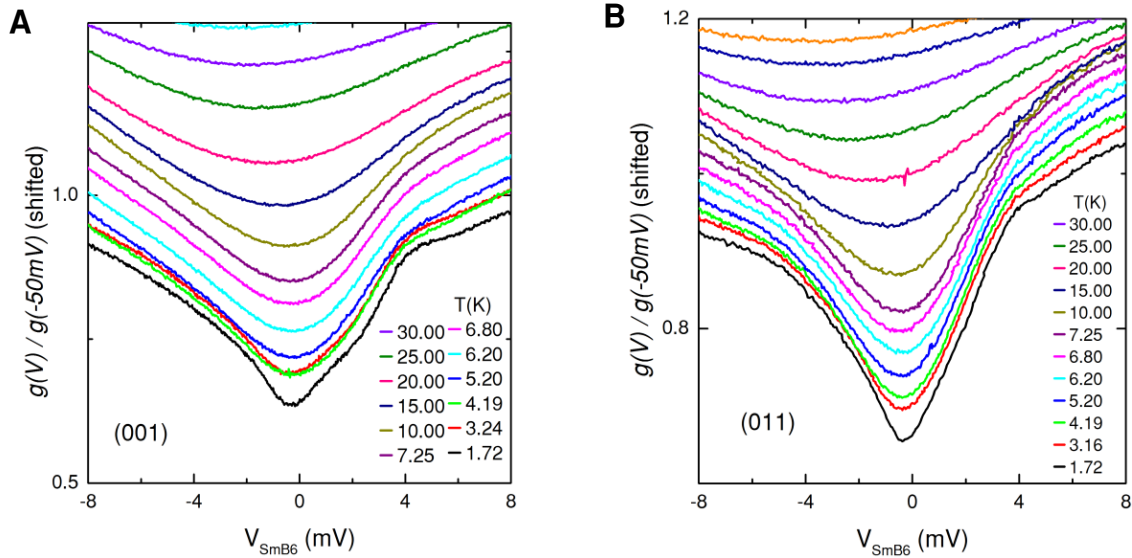

**Fig. S6.** Temperature evolution of the kink and hump structures. Junctions on (*A*) the (001) surface and (*B*) the (011) surface. All curves except the 1.72 K one are shifted vertically for clarity. The kink and hump structures seem to show up below 15 K – 20 K, in agreement with the range where the decreasing behavior in *g(V_b)/g(–50mV)* turns into an increasing one and subsequent saturation at low temperature (see Fig. 1*D* & Fig. S5).

## 5. Pb Phonon Features in Tunneling

The temperature evolution of conductance below the $T_c$ of Pb is shown in Fig. 3A. The $V_1$ peak evolves continuously (in both position and height) to turn into a kink above the $T_c$. This indicates that their origins are the same, namely, interaction with spin excitons, as discussed in the text. As argued in the main text, this peak is not the phonon feature typically seen in tunnel junctions on Pb (10) for several reasons as discussed in the main text. For instance, as shown in Fig. S7, such phonon features should appear symmetrically and its shape is not a peak but a hump-dip. Moreover, with increasing temperature, they become smeared out more quickly than the coherence peaks, whereas the $V_1$ peak remains more pronounced than the coherence peaks, again, indicating that it is not due to the phonons in Pb.

An early report (11) based on neutron scattering measurements on $SmB_6$ shows phonon modes at 2.6 and 11.6 meV, different from the locations of the $V_1$ peak and the kink. This indicates that the phonons in $SmB_6$ are not responsible for these features, either.

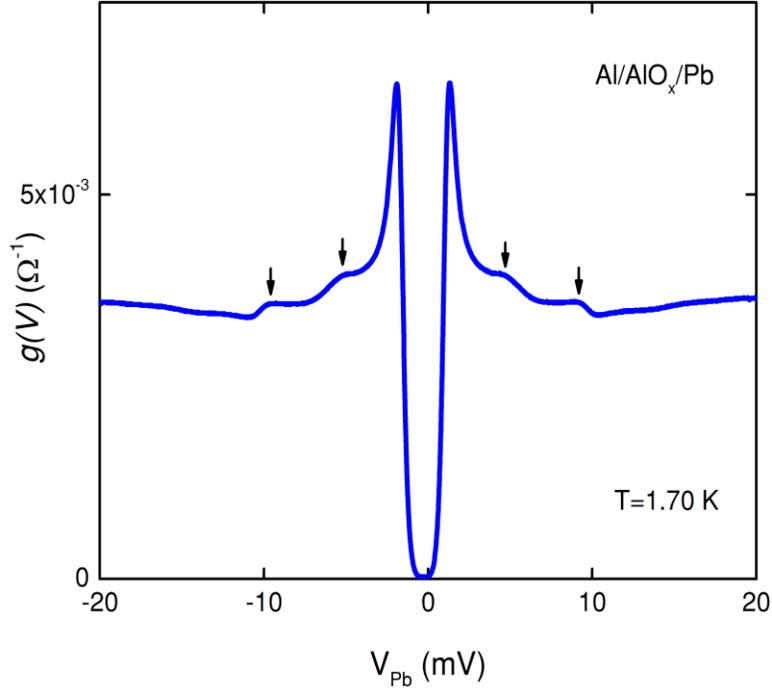

**Fig. S7.** Pb phonon features in tunneling. Example tunneling data from an Al/AlO$_x$/Pb tunnel junction. The arrows point to the hump-dip structures in the superconducting DOS of Pb due to strong coupling to the phonons (transverse & longitudinal acoustic phonons at ~4.5 mV and ~8.5 mV, respectively). They appear symmetrically as do the coherence peaks since they are embedded in the DOS (bulk physics).

## 6. Reproducibility of Conductance Features

The characteristic features in tunneling conductance reported in the main text are reproducibly observed in junctions prepared following the optimized procedure described in Sect. 1. Figure S8 shows a set of such data taken at low temperatures. The bulk gap features including asymmetric background shape and the peak at −21 mV are reproduced. Features due to the topological surface states are repeatedly seen: double vs. single Dirac cone(s) and kink-hump structures. Locations of Dirac points and the kinks are slightly different from the ones presented in the main text, presumably due to slight differences in the chemical potential and the length scale for interaction with spin excitons at the surface depending on the condition of surface oxidization. Also, the peak behaviors below the $T_c$ of Pb are reproducible. The $V_{1(2)}$ peak is located at a bit lower bias than in the text, consistent with the concomitantly smaller value for $\omega_{1(2)}$. This further supports the interpretation based on inelastic tunneling involving spin excitons.

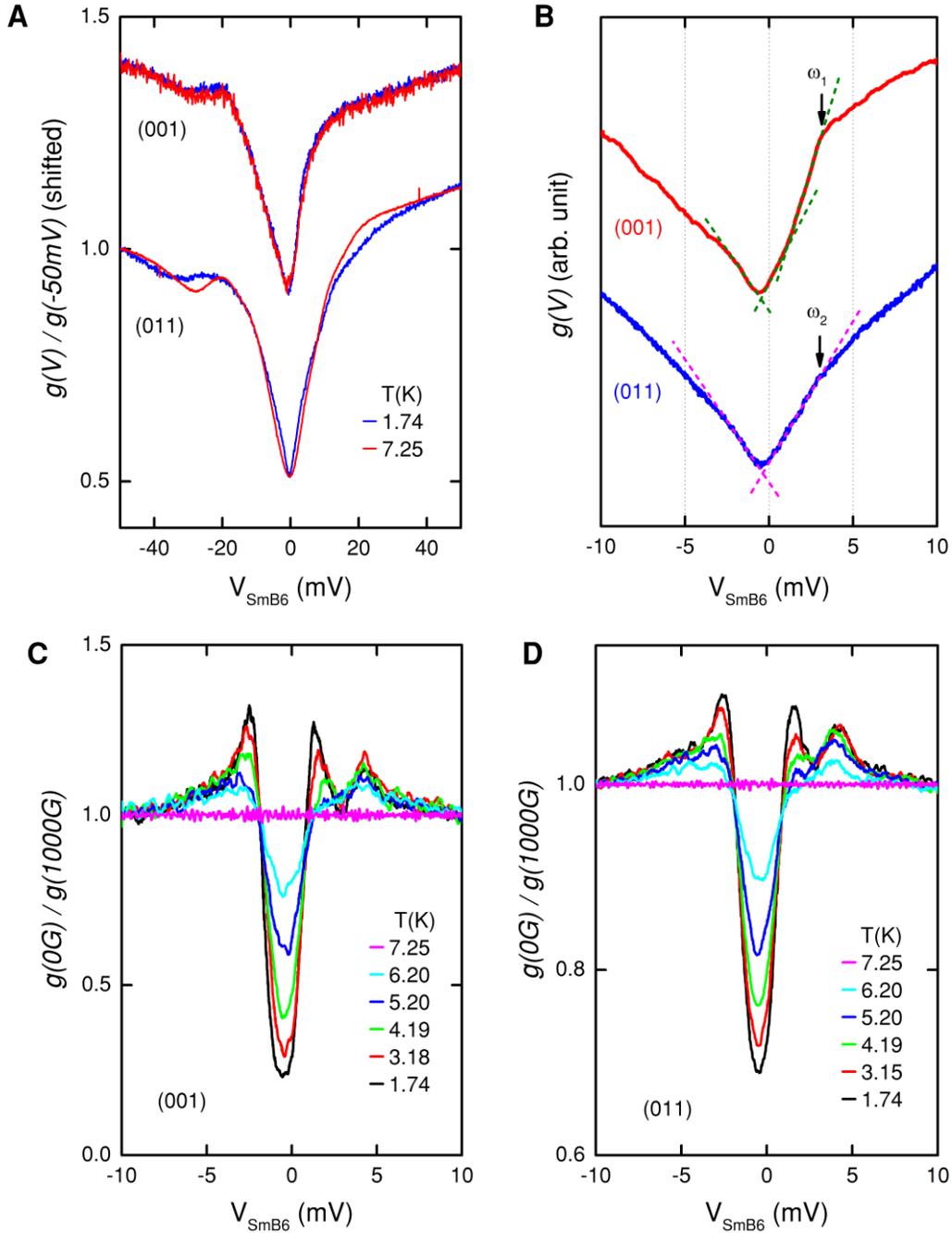

**Fig. S8.** Reproducibility of conductance features. (*A*) Overall conductance shape showing the bulk gap features. Here Pb is in the normal state. Curves are shifted vertically for clarity. (*B*) Linear conductance in the low bias region arising from double vs. single Dirac cones, respectively. (*C* and *D*) Conductance below the $T_c$ of Pb normalized similarly as in Fig. 3*B*. The asymmetric appearance and temperature evolution of the $V_{1(2)}$ peak as well as the relative height of the V− peak are similar to those reported in the main text. These features are reproducibly observed in all working tunnel junctions, indicating their robust origin.

## 7. Hybridization Gap in SmB$_6$

The tunneling conductance into a heavy fermion Kondo lattice has been formulated theoretically by several groups (12-14). Such tunneling models have been adopted to analyze conductance data taken from heavy fermion Kondo lattices and Kondo insulators (15-17). Based on the idea of co-tunneling, Maltseva *et al.* (12) derived an explicit expression as follows.

$$\left.\frac{dI}{dV}\right|_{FR} \propto \mathrm{Im}\,\tilde{G}_\psi^{KL}(eV);\ \tilde{G}_\psi^{KL}(eV) = \left(1 + \frac{q_F W}{eV - \lambda}\right)^2 \ln\left[\frac{eV + D_1 - \frac{v^2}{eV - \lambda}}{eV - D_2 - \frac{v^2}{eV - \lambda}}\right] + \frac{2D\left(\frac{t_f}{t_c}\right)^2}{eV - \lambda}, \quad (S1)$$

where $\lambda$ is the renormalized $f$-level, $v = z^{1/2} v_0$ is the renormalized hybridization matrix amplitude with $z = 1 - n_f$ ($n_f$: $f$-level occupancy), $-D_1$ and $D_2$ are the lower and upper conduction band edges, respectively. The Fano parameter $q_F = t_f v / t_c W$, where $t_f$ and $t_c$ are the tunneling matrix amplitudes for the $f$-orbital and the conduction band, respectively. The indirect gap is given as $\Delta_{\mathrm{hyb}} = 2v^2/D$ ($2D$: conduction bandwidth).

Attempts have been made to analyze the experimental data using this model. Figure S9 shows such a computed conductance curve. As shown, it is found that the data cannot be fitted quantitatively, which is because there are many additional features not accounted for by this model. The extracted hybridization gap size is about 21 meV, quite different from an inferred value (~30 meV) based on the conductance shape. This clearly shows that a realistic gap size should be obtained from a quantitative analysis, not via eyeballing. In the end, it is desirable to develop an extended model for tunneling into a topological Kondo insulator to include the spectral density of the surface states influenced by the interaction with spin excitons. It is more fundamental to develop formulations for the description of microscopic tunneling process into such a complex system. Here, it will be necessary to identify relevant length scales (5, 18-20) and understand their temperature evolution.

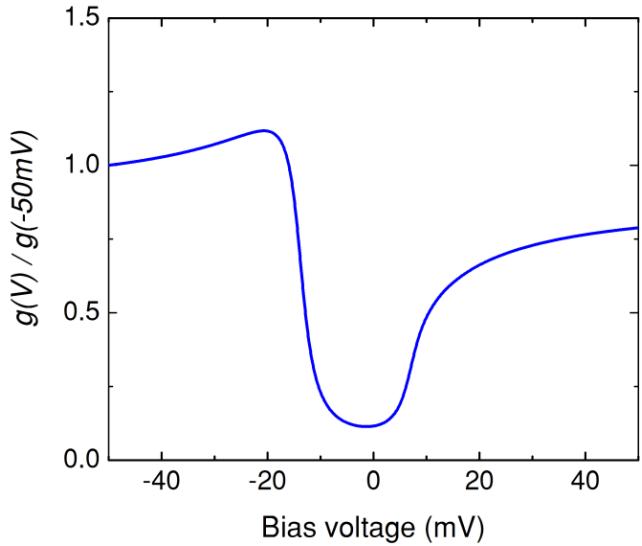

**Fig. S9.** Simulated conductance curve. The model by Maltseva *et al.* (12) is adopted to replicate the experimentally observed conductance shape for the bulk SmB$_6$ including the background asymmetry and the peak position at −21 mV. Here, used and derived parameter values are: $T = 1.72$ K, $T_K = 50$ K, $D_1 = D_2 = 2.3$ eV, $W = 1\ k_B T_K$, $\lambda = -3.4$ meV, $q_F = -0.7$, $v = 155.1$ meV, $t_f/t_c = -0.0194$, and $\Delta_{\mathrm{hyb}} = 20.9$ meV. A constant value of $\gamma = 0.5\ k_B T_K$ is used for the quasiparticle broadening parameter.

## 8. Analysis of the Linear Conductance Due to Topological Surface States

As is well known, tunneling conductance for a simple system where no interference effect such as Fano resonance (21) exists is directly proportional to the DOS. A good example is the case of tunneling into a superconductor. However, in a non-superconducting system without interaction (simple metal), the DOS factor cancels out with the Fermi velocity in the conductance kernel, Harrison's theorem (22), so the DOS is not detectable by tunneling spectroscopy unlike the case of superconductors.

For Dirac fermions following a linear dispersion, $E = \hbar v_F k$, the DOS is expressed as $N(E) = 2A_c|E|/\pi\hbar^2 v_F^2$, where $A_c$ is the area of the unit cell (23). In this case, apparently, the DOS doesn't cancel out with the Fermi velocity. Thus, one can expect to observe the DOS in the tunneling spectroscopy as is demonstrated by numerous measurements on Dirac fermion systems including graphene and topological insulators. Based on our observation of the linear conductance in the low-bias region, we can express the DOS for the surface states as follows:

$$N_i(E) = \frac{2A_{ci}}{\pi\hbar^2 v_F^2}\left|E - E_D^i\right|, \tag{S2}$$

where $E_D^i$ is the Dirac point for the band $i$. That is, the linear dispersion for topological states gives rise to a V-shaped DOS, which is observed as linear conductance around certain bias voltages corresponding to Dirac points.

Assuming the tunneling matrix is independent of energy, the tunneling current across a barrier in a planar junction is expressed as follows:

$$I(V) = S|T|^2 e\int_{-\infty}^{+\infty} N_t(E) N_b(E+eV)[f(E) - f(E+eV)]dE \tag{S3}$$

where $S$ is the junction area, $T$ is the tunneling matrix element, $f$ is the Fermi-Dirac distribution function, and $N_t$ and $N_b$ are the density of states of the top and bottom electrode, respectively. The differential conductance can be found by differentiating the current, that is, $g(V) = dI/dV$. Thus, without knowing the tunneling matrix element, it is not feasible to extract an absolute value for the Fermi velocity. However, by comparing the slopes of the decomposed linear conductance, which are inversely proportional to $v_F^2$, and comparing them with the Fermi velocities extracted in quantum oscillation measurements (5), we can still identify the symmetry points around which each Dirac cone resides. For the (001) surface, the ratio of the slopes is found to be 3, giving rise to 1.74 for the ratio of the Fermi velocities. This Fermi velocity ratio is not too much different from 2.24, a value obtained in quantum oscillation measurements (5), so we infer that one Dirac cone is around the $\bar{\Gamma}$ point with smaller $v_F$ ($\alpha$ band) and the other one is around the $\bar{X}$ point with larger $v_F$ ($\gamma$ band). The slope in the (011) data is comparable to the larger one in (001) but this comparison is not accurate since the pre-factors in the expression for the DOS and the tunneling conductance would be different among different junctions and measurements. The Fermi velocity for this surface state ($\beta$ band around $\bar{Y}$ point) is reported to be in between those for the $\alpha$ and $\gamma$ bands. This is the case when the slopes are compared by taking the pre-factors into account, but the ratios of $v_F$ don't match exactly with the quantum oscillation results.

## 9. Surface Spectral Density Influenced by Spin Excitons

Following the theory by Kapilevich *et al*. (18), the surface spectral density is calculated. The self-energies used in this calculation are to account for the interaction of surface Dirac fermions with spin excitons. Figure S10 shows an example in which the DOS due to two Dirac cones is adopted to replicate the double-linear conductance observed in the (001) surface. The two Dirac points used are at −0.4 meV and 3.0 meV. While this is enough to illustrate the influence of such interactions, a full-blown model is in need to fit the data quantitatively using this theory since the surface states conductance appears to be superimposed on top of the gapped DOS of the bulk.

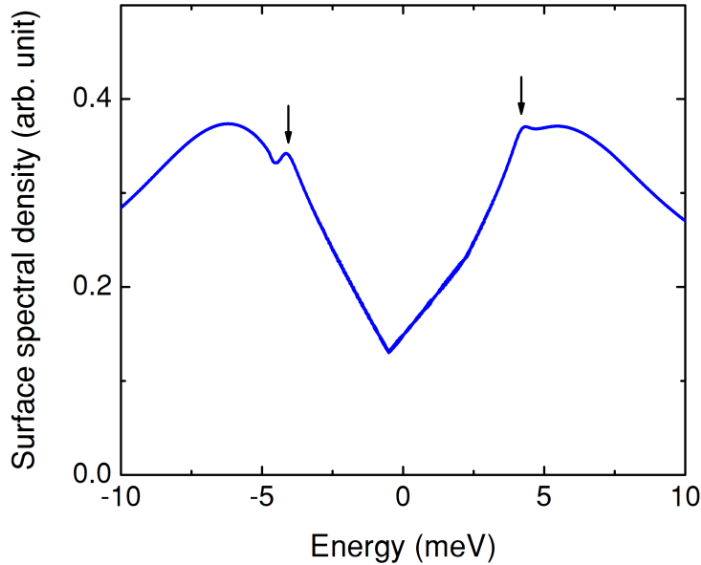

**Fig. S10.** Simulated spectral density for the topological surface states interacting with spin excitons. Here, non-interacting DOS for two Dirac cones is used to replicate the conductance shape seen in the (001) surface (Fig. 2*A*). The two Dirac points are at −0.4 meV and 3.0 meV. The temperature is set to 1.72 K. The arrows indicate peaks caused by the interaction. The linearity is lost outside the peaks, indicating that the interaction destroys the topological protection.

## 10. Comparison with ARPES Results

There are several reports based on ARPES measurements in $SmB_6$ (24-31). It is interesting to note that one of them (25) presents intensity vs. energy plots showing features similar to those seen in our tunneling conductance data. Figure S11 is to compare these two different measurements. Both measurements show a peak due to the bulk gap at a similar energy scale, around −20 meV. The authors of Ref. 25 interpret the hump appearing near the Fermi level as a signature for the in-gap states. Considering the similarity in its shape, energy scale, and temperature evolution to those in our conductance spectra, we believe that it originates from the same source, namely, interaction of surface states with spin excitons. If the ARPES measurement had been carried out at as low temperature as in our tunneling spectroscopy, the intensity would have shown a linear region close to the Fermi level as seen in tunneling.

    There are several subtle differences between the tunneling and APRES results, caused by the differences in the two techniques. First, ARPES detects mostly filled states since a different instrumentation is needed to detect empty states via inverse photoemission process, whereas tunneling can probe both filled and empty states using the same junction and, thus, can reveal the kink at a positive bias. Second, in addition to the higher energy resolution, planar tunneling spectroscopy provides another advantage that it can probe a system using different counter-

electrodes (normal metal or superconducting) enabling investigation of detailed behaviors as reported in this work. Third, of course, ARPES is a momentum-resolving technique whereas planar tunneling doesn't have such momentum resolution but momentum selectivity. Therefore, owing to its momentum resolving power, ARPES enables to measure the spectral density from each of the two Dirac cones separately as seen in Fig. S11. On the other hand, our tunneling conductance from nominally the same (001) surface reveals the two Dirac cones via the double linear conductance as it is the sum of contributions from the two Dirac cones.

In any case, it is remarkable that these two different techniques can provide spectroscopic information for the same underlying physics in $SmB_6$. After all, if everything is right, different techniques should provide consistent information on the same physics. This observation is in contrast to the case of scanning tunneling spectroscopy (STS), which has not shown much similarity to our tunneling results. More in-depth discussion of this point is given in Sect. 11.

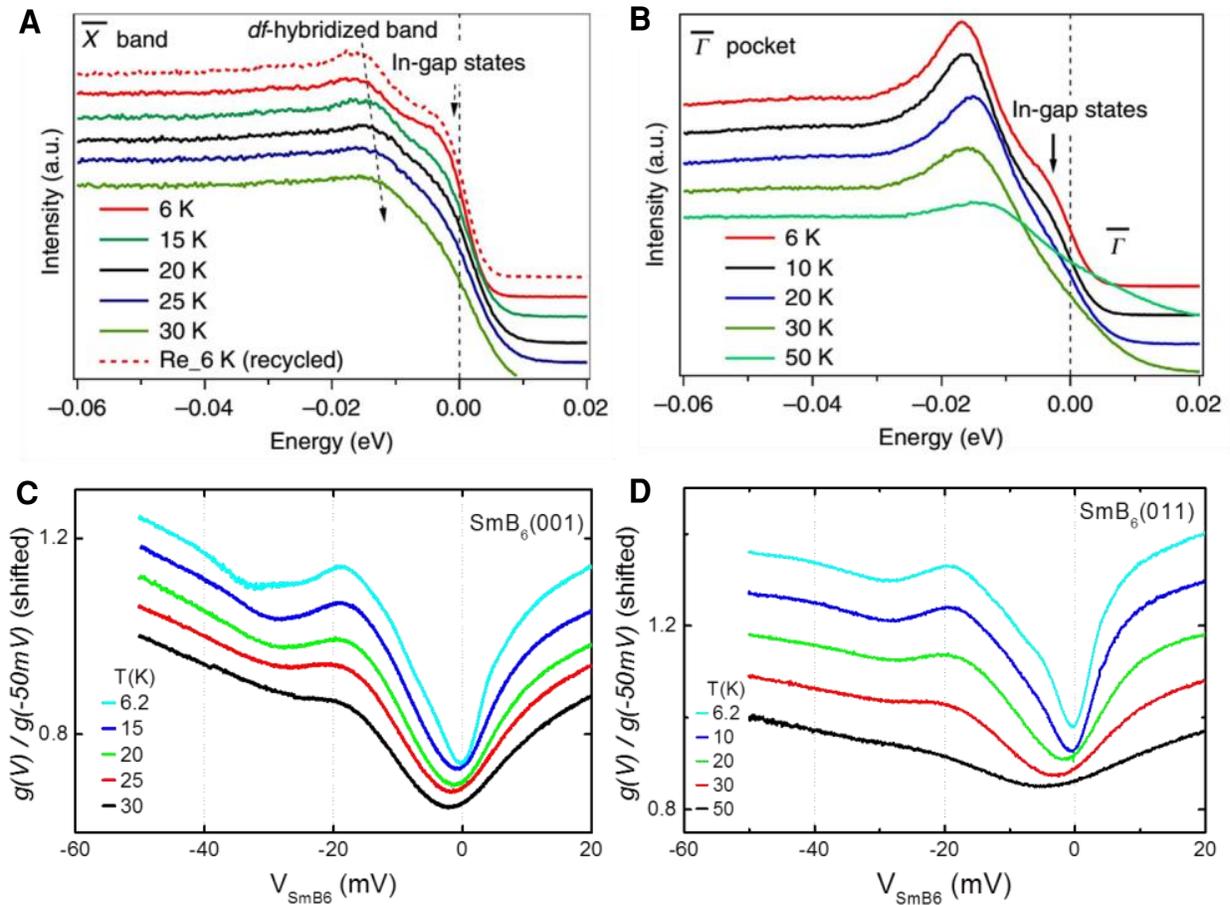

**Fig. S11.** Comparison between ARPES and planar tunneling. (*A* and *B*) ARPES data on a (001) surface, copied from Ref. 25 with permission from the publisher. (*C* and *D*) Our tunneling conductance spectra plotted over the same energy scales and temperature ranges as in the ARPES data for comparison purpose. Notice the similarity in the bulk gap feature (peak at −20 mV) and the hump structure slightly below the Fermi level, which we interpret as due to the interaction with spin excitons.

## 11. Inelastic Tunneling Spectroscopy

A tunnel junction consists of two electrodes separated by a tunnel barrier. Quantum mechanical tunneling of quasiparticles can occur via an elastic or inelastic process. For simplicity, let us consider cases that don't involve interference effects such as Fano interference (21). Then, if tunneling occurs only elastically, the tunneling conductance can directly map out the DOS of the electrodes provided that the Harrion's theorem (22) (also see Sect. 8) doesn't hold. A normal-metal/insulator/superconductor tunnel junction is such an example (see Fig. S7). If inelastic channels are added as shown in Fig. S12*A*, the tunneling current and, thus, the conductance will be enhanced above eV = $\omega_0$, where $\omega_0$ is the energy required for such inelastic process (Figs. S12 *B* & *C*). As is well known, the source of inelastic channels is best identified by a pronounced peak in the second harmonics measurement (Fig. S12*D*). If one of the electrodes is a weakly-correlated topological insulator like $Bi_2Se_3$, the differential conductance will be similar to a simple metallic case except for the linear conductance due to Dirac fermions' V-shaped DOS as illustrated in Fig. S12*E*. Also, the conductance should be symmetric with respect to the bias voltage. However, if the topological insulator contains excitations which (i) are the only sources for inelastic tunneling channels and (ii) strongly interact with the Dirac fermions, the conductance features will look quite different. This is the case for $SmB_6$. As shown in Fig. S12*F*, the linearity stops at eV = $\omega_0$ because of the interaction with spin excitons. The conductance shape is asymmetric since the conditions for emission and absorption process are different as discussed in the main text and as clearly seen in our data.

As noted earlier, the kink energy appears well-defined in each conductance curve. Contrary to this, the hump in the negative bias region is quite broad, so no characteristic energy scale can be assigned to it but just a range. These distinct behaviors can be understood by considering the asymmetric process for emission vs. absorption. The kink appears at $\omega_0$ since this is the threshold energy for emission of spin excitons set by the bulk physics of $SmB_6$, namely, collective excitations. Meanwhile, the absorption process can occur at any negative bias as long as spin excitons are available since electrons at energy $\omega_0$ below the chemical potential can pop up into the chemical potential by absorbing spin excitons. Due to this background contribution, this process would results in only a weak maximum in conductance around eV = $-\omega_0$. After all, this is a kind of resonance process. The reason why it shows up as a broad hump instead of a peak is because this perturbative contribution to the spectral density is added to the linear (instead of flat) background DOS of the surface states in $SmB_6$.

The tunneling model based on inelastic processes in $SmB_6$ as presented in the paper can explain the experimental features quite consistently. Whether such model involving the phonon emission and absorption in Pb can also work reasonably, diagrams corresponding to the specific bias configurations are compared in Fig. S13. Apparently, this scenario cannot properly explain the asymmetric features of the $V_{1(2)}$ and V– peaks.

One of the questions that may arise regarding the inelastic tunneling described in the text is whether the features observed in our planar tunneling spectroscopy can also be seen in STS (17, 32, 33). In simple cases such as illustrated in Fig. S12A, it is true that both techniques have been successful to identify such inelastic channels (34). In such cases, an apparent advantage of STS is that one can resolve the spatial dependence by moving a tip close to local species responsible for inelastic processes. There have not been many reports when the inelastic process involves *collective excitations* in the electrode itself. One of the rare examples is the phonon-involved inelastic tunneling in graphene flakes measured with an STM (35). It is an open question whether a similar STS measurement is possible on $SmB_6$ to detect the spin-exciton features as observed in our planar tunneling spectroscopy. Although the same fundamental quantum mechanical process is exploited in both techniques, several factors should be carefully considered including probing depth, momentum selectivity, and local vs. global nature (broken (14) or unbroken translational symmetry). For instance, is the tunneling current in STS large enough to excite spin excitons that occur collectively in $SmB_6$? If the challenge has to do with the surface chemistry, it may not be possible to detect those features using an STM without pacifying the dangling bonds at the surface as might be done favorably via the surface oxidation in our junction fabrication procedure.

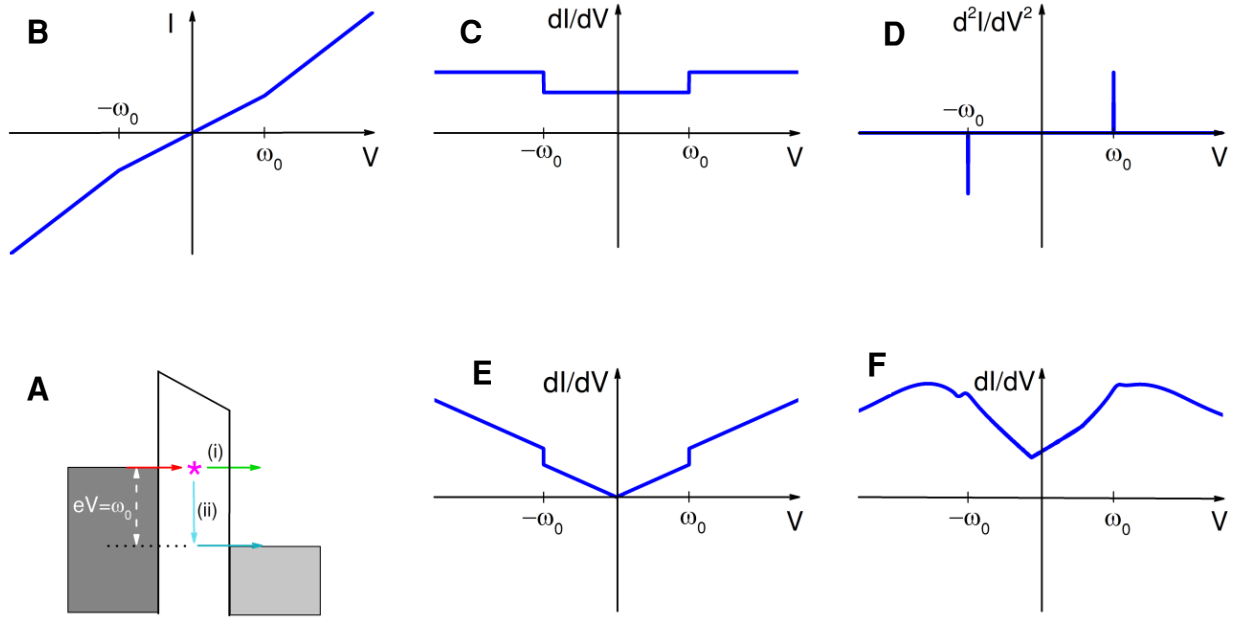

**Fig. S12.** Inelastic tunneling spectroscopy. (*A*) Schematic of a tunnel junction. The asterisk in the barrier denotes a source for an inelastic tunneling channel such as molecules and atomic chains. The process labeled as (i) depicts elastic tunneling, whereas (ii) is for inelastic tunneling which can occur only when the energy of tunneling electrons (eV) is equal to or greater than the energy ($\omega_0$) required for the inelastic process. (*B*, *C* and *D*) Characteristics of current, differential conductance, and second derivative vs. bias voltage, respectively, for inelastic tunneling as depicted in *A* with simple metallic electrodes. (*E*) Differential conductance when one of the electrodes is a topological insulator such as $Bi_2Se_3$. (*F*) Differential conductance when the electrode is $SmB_6$ where the inelastic process (emission or absorption of spin excitons) occurs, not in the barrier, as discussed in the main text and in Sect. 9.

## 12. Relevant Length Scales

As mentioned in the main text, it remains to be investigated further how the planar tunneling spectroscopy works in such complicated junction structures involving $SmB_6$. More specifically, (i) How are the bulk gap features detected in this surface sensitive measurement? (ii) What are the relevant length scales such as the Fermi wavelength ($\lambda_F$) and diffusion length ($l_s$) of the surface states? Using the results from quantum oscillation measurements (5), $\lambda_F$ is estimated to fall to the range of several nm, which is longer than in simple metals by more than one order of magnitude. Some of the recent theoretical works (18-20) have estimated $l_s$ to be several lattice spacing. Thus, one may speculate that the relevant length scales are long enough for the planar tunneling to probe both the surface states and the bulk gap. Since the antiferromagnetic coupling *J'* is expected to decrease as moving toward the surface (18), the spin exciton energy is also expected to decay toward the surface. Therefore, regarding the impact of interaction with spin excitons, the length scale $l_s$ relative to this decay length of spin exciton energy would play an important role.

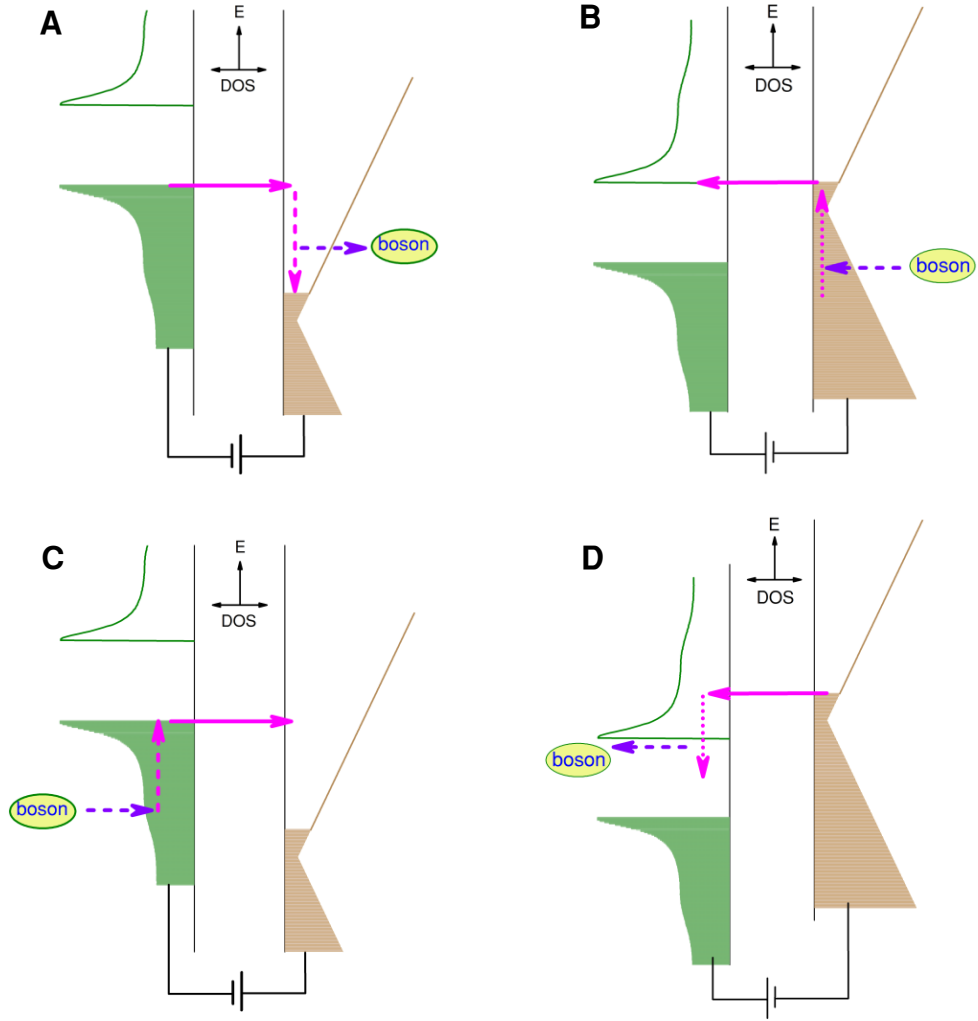

**Fig. S13.** DOS diagrams for the specific bias configurations described in the text. (*A* and *B*) Inelastic tunneling involving bosons (spin excitons) in $SmB_6$. (*C* and *D*) Inelastic tunneling involving bosons (phonons) in Pb. The bias for *A* and *C* is $eV = \Delta + \omega_0$. Since the phonon absorption process depicted in *C* can occur at any bias, it will only add a constant contribution instead of the pronounced $V_{1(2)}$ peak as seen in the data. The phonon emission process depicted in *D* cannot explain the V– peak at $eV = -\Delta$ since such process requires tunneling electrons to arrive with minimum excess energy of $\omega_{ph}$.